\documentclass[pra,aps,superscriptaddress,twocolumn,floatfix,notitlepage,longbibliography]{revtex4-2}

\usepackage{amsmath}
\usepackage{amsfonts}
\usepackage{amssymb}
\usepackage{graphicx}
\usepackage{color}
\usepackage{xcolor}
\usepackage{xspace}
\usepackage{bbold}
\usepackage{siunitx}
\usepackage{float}

\usepackage[sort&compress]{natbib}

\newcommand{\tG}{t_\mathrm{G}}

\usepackage[colorlinks, linkcolor=magenta, citecolor=magenta,
urlcolor=blue]{hyperref}
\begin{document} 
%\title{Photon correlations of ideal single-photon sources}
%\title{The relation between continuous-wave and pulsed single-photon
%  sources}
%\title{Connecting continuous-wave and pulsed single-photon sources}
%\title{The relation between continuous-wave and pulsed single-photon sources}
%\title{The relation between continuous-wave and pulsed single-photon emission}
\title{Perfect single-photon sources}

\author{Sana Khalid}
\affiliation{Faculty of Science and Engineering, University of Wolverhampton, Wulfruna St, Wolverhampton WV1 1LY, UK}
\author{Fabrice P. Laussy} 
\affiliation{Faculty of Science and Engineering, University of Wolverhampton, Wulfruna St, Wolverhampton WV1 1LY, UK}

\date{\today}

\begin{abstract}
  We introduce the \emph{gapped coherent state} in the form of a
  single-photon source (SPS) that consists of uncorrelated photons as
  a background, except that we demand that no two photons can be
  closer in time than a time gap~$\tG$. While no obvious quantum
  mechanism is yet identified to produce exactly such a photon stream,
  a numerical simulation is easily achieved by first generating an
  uncorrelated (Poissonian) signal and then for each photon in the
  list, either adding such a time gap or removing all successive
  photons that are closer in time from any photon that is kept
  than~$\tG$. We study the statistical properties of such a
  hypothetical signal, which exhibits counter-intuitive features. This
  provides a neat and natural connection between continuous-wave
  (stationary) and pulsed single-photon sources, with also a bearing
  on what it means for such sources to be perfect in terms of
  single-photon emission.
\end{abstract}

\maketitle

\section{Introduction}

The single-photon source (SPS) is a cornerstone in photonics and for
quantum-optical
technology~\cite{obrien09a,senellart17a,couteau23a,couteau23b}. It
describes light that is produced photon by photon, thereby fighting
their urge (as bosons) to pile-up together, as they do when left in
thermal equilibrium~\cite{hanburybrown56c}.  Since Glauber's theory of
photon correlations~\cite{glauber63a}, the quantity of choice for the
SPS is the second-order coherence function $g^{(2)}(\tau)$ that
quantifies the density of two-photons separated in time by~$\tau$. A
SPS should in particular be such that no two photons are ever found
together, so $g^{(2)}(0)=0$ is wanted. In practice, though,
$g^{(2)}(0)$ is ``only'' much smaller than unity, with $g^{(2)}=1$
describing uncorrelated photons, whose proximity is left to
chance. The failure to be exactly zero allows for cases where
$g^{(2)}(0)\ll 1$ while $g^{(3)}(0)\gg 1$, the latter quantity
describing three-photon coincidences (this could also be true for
any~$g^{(n)}$). In contrast, if~$g^{(n)}(0)=0$ (exactly) then all
$g^{(m)}(0)$ with~$m\ge n$ must also be zero. While it certainly may
seem strange that a source which suppresses strongly its two-photon
coincidences does not compulsorily also suppress higher-order ones,
this is nevertheless a possibility (as long as $g^{(2)}(0)$ is not
\emph{exactly} zero). This illustrates how subtle photon correlations
can be~\cite{zubizarretacasalengua17a,grunwald19a} and how difficult
it is to define a good criterion for
SPS~\cite{arXiv_lopezcarreno16c}. In this text, we consider
hypothetical signals that will allow us to precise such notions.

\section{The concept of perfect single-photon emission}

It is clear from our introduction that $g^{(2)}(0)=0$ \emph{exactly},
is a requisite for the perfect single-photon source. What would such
an ideal source look like, if we could further conceptualize its
attributes?  There, the dynamics of emission and detection enters
the picture. Indeed, one could imagine a steady-state SPS where, at
each and every moment of time, the field has a single photon. This
would constitute a continuous (all-times) yet still quantized
(one-photon) field. This is a strange, ultimate quantum field. The
emission of an incoherently-pumped two-level system quenched in its
excited state approaches such a scenario in the limit of infinite
pumping rate: the system is constantly kept excited as it constantly
emits single-photons. This is a highly-pathological object which, for
intance, is impossible to simulate as a point process with ``clicks''
by, say, quantum Monte Carlo procedure~\cite{molmer93a}, since the
time steps should be vanishing. This makes obvious the fundamental
problem that one should undertake an exact measurement in time to
observe only a single photon from such a source. The slightest
holding-up of time will accumulate several of the single photons and
spoil its SPS character. A manifestation of such a pathological
feature is that the luminescence spectrum of such a source is flat
(photons are emitted at all the frequencies) with a vanishing height,
as the area is unity, corresponding to the population of the emitter
(which is maintained in its excited state). Nevertheless, this
continuous flow of photons is such that $g^{(2)}(0)=0$, exactly. In
this case, the emissions at all, i.e., including infinitely large
frequencies, correspond to the vanishing times of the single-photon
emissions.

\begin{figure}[tb]
  \includegraphics[width=.9\linewidth]{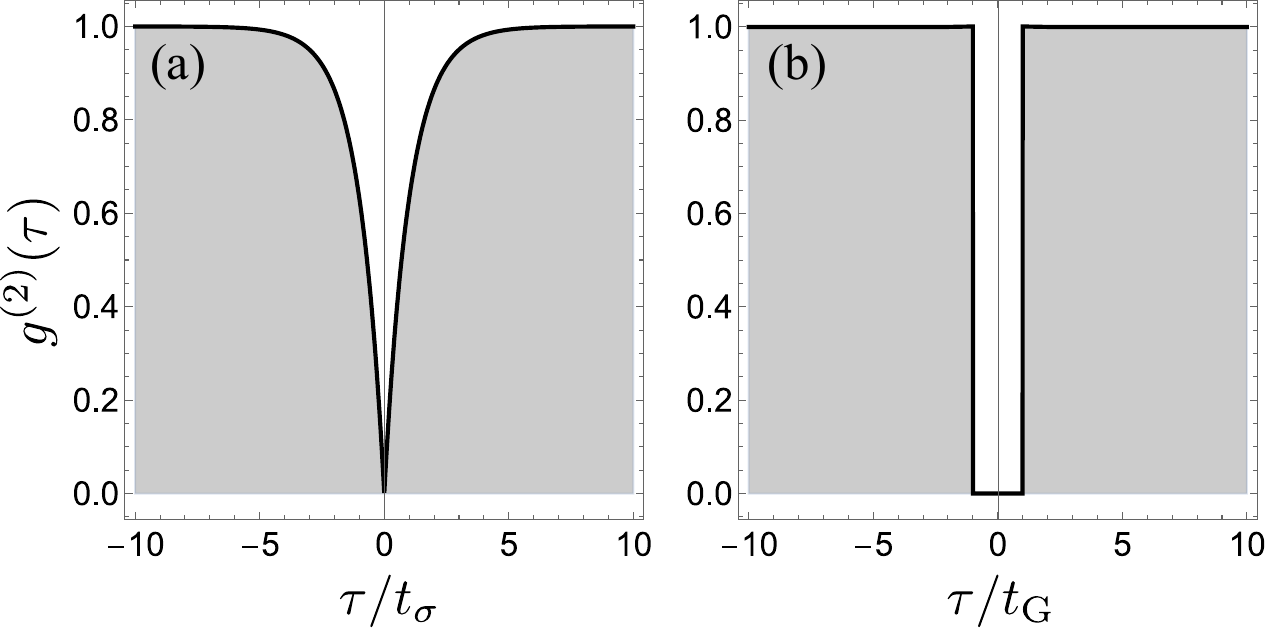}
  \caption{Glauber's second-order correlation function $g^{(2)}(\tau)$
    for (a) the incoherently-driven two-level system, in units of
    $t_\sigma=1/\gamma_\sigma$ and (b) a gapped single-photon source,
    with vanishing~$\gamma\tG$.}
  \label{fig:Wed24May130615BST2023}
\end{figure}

It is thus straightforward to have a dynamical system that results in
$g^{(2)}(0)=0$ exactly, but this comes as an unphysical limit for the
optimum emission rate: the ideal perfect single-photon source is
pathological (here note that we use two similar adjectives to refer to
different attributes). Turning to finite pumping rate~$P_\sigma$, the
emission of a two-level system still features exactly $g^{(2)}(0)=0$
and the spectrum becomes physical: a Lorentzian of width
$\gamma_\sigma+P_\sigma$ and area
$P_\sigma/(\gamma_\sigma+P_\sigma)$. We have insisted on the extreme,
unphysical case above as it actually captures the key problem in
realizing a single-photon source: the perfect time-resolution for the
detection is still needed also for the physical SPS. Although an
incoherently-pumped two-level system spaces its successive emitted
photons in time, and despite the suppression indeed of their proximity
as compared to a Poisson process, detection in time will always
accidentally collect various photons together, unless there is no time
uncertainty whatsoever, i.e., the detection is temporally
point-like. This brings forward the problem of the photon correlations
in time, through $g^{(2)}(\tau)$ where~$\tau$ is the time-delay
between any two detected photons. For a two-level system,
$g^{(2)}(\tau)=1-\exp(-\gamma|\tau|)$ as shown in
Fig.~\ref{fig:Wed24May130615BST2023}(a). The theory of
photo-detection~\cite{delvalle12a} shows that any measurement comes
with a finite time-resolution~\cite{lopezcarreno22a} and since
$g^{(2)}(\tau)=0$ exactly, only for~$\tau=0$ exactly, any overlap with
nonzero times bring us back to an imperfect $g^{(2)}(0)$.  In frequency
space, this means that although emitting with a physical Lorentzian
spectrum, one will not get perfect antibunching unless one detects all
the frequencies, which is in practice impossible, since Lorentzian
have fat tails and so, regardless of the bandwidth of the detector,
some photons will always escape and, ironically, this results in
bunching~\cite{delvalle12a}. Such fundamental aspects tend to be
underappreciated and relegated to technical limitations from the
setup, instruments and/or the samples. While those certainly still
take the largest share of the problem, even if they could be mitigated
or even suppressed altogether, one would still face the fundamental
problem we have just discussed.  Now that we have highlighted the
situation with paradigmatic two-level SPS (this is further discussed
in Refs.~\cite{arXiv_lopezcarreno16c,lopezcarreno22a}), we can cut the
Gordian knot and turn to a scheme that would, in principle and at a
fundamental level, provide a perfect SPS.

\begin{figure*}
  \includegraphics[width=.75\linewidth]{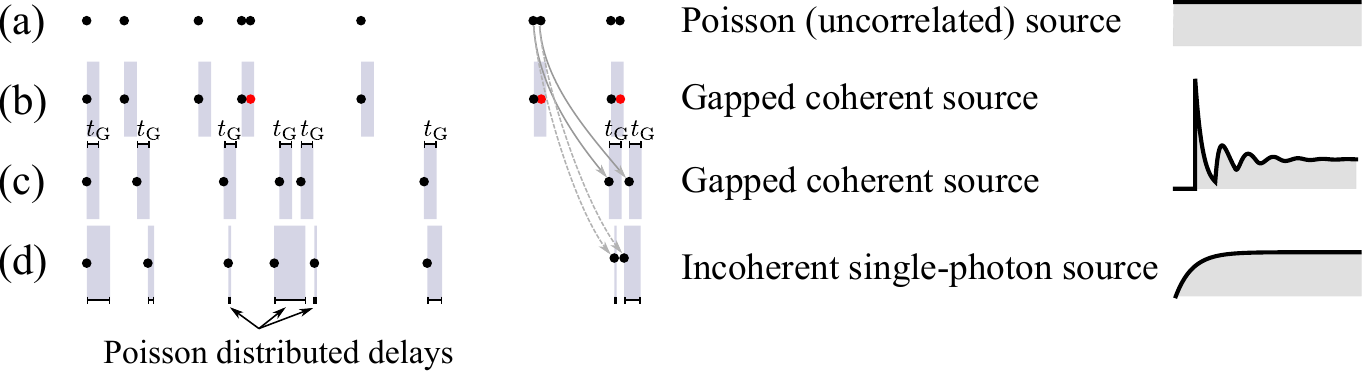}
  \caption{A point process view of photon emission: in~(a), a typical
    sampling of ten points randomly distributed in time, with no
    correlations, describing photons from a coherent states. In~(b)
    and~(c), two possible constructions are shown of a gapped coherent
    source from the backbone Poisson stream, formed by (b)~removing
    photons closer to their predecessor than the time gap~$\tG$, here
    stripping off the red points, or by~(c) adding the time gap as an
    extra ``spacer'' between every photons from the original
    stream. If such a repulsion is not rigid but also randomly
    distributed according to a Poisson distribution, as shown in~(d),
    instead of a gapped coherent source, one recovers the emission
    from a conventional incoherently pumped two-level system.
    Although infrequent, Poisson repulsion can result in some photons
    bunching, due to the Poisson bursts, as shown for the 7th and 8th
    photons in stream~(a), that may remain very close to each other as
    shown in~(d), in contrast to their gapped counterpart~(c).}
  \label{fig:Wed24May120508BST2023}
\end{figure*}

\section{Gapped coherent state}

In this text, we do not consider quantum-dynamical mechanisms to
produce light but directly streams of photons as events in time (time
series) and study their general mathematical structures following
various desiderata that one might require from a SPS (from now onward,
it will therefore be convenient to undertand SPS as ``single-photon
stream'')~\cite{khalid_thesis24a}.  In particular, this allows us to
consider what, following our previous discussion, constitutes clearly
a ``perfect SPS'' even in the face of imperfect detectors: one needs
to open a gap in time between successive photons, to enforce, by
construction, no-coincidences not only at~$\tau=0$ but at all times
within the gap, i.e.,
\begin{equation}
  \label{eq:Sat22Apr190318BST2023}
  g^{(2)}(\tau<\tG)=0\,.
\end{equation}
Outside of the gap, we may expect photons to have no correlations and
thus
\begin{equation}
  \label{eq:Sat22Apr190411BST2023}
  g^{(2)}(\tau\ge \tG)=1
\end{equation}
as shown in Fig.~\ref{fig:Wed24May130615BST2023}(b). This can be
obtained from a Poisson background as the simplest case---which also
corresponds to the popular coherent state---by opening a time gap in
its emission, and thus we refer to this hypothetical signal as the
``\emph{gapped coherent state}''.  For such a stream of photons, no
two photons are ever found in the same time interval~$\tG$, regardless
of where this is sampled. One possible way to construct it as a
pointlike process is, starting from a random (Poisson) sequence
(Fig.~\ref{fig:Wed24May120508BST2023}(a)), to remove every instance
that, if leaving it there, would result in more than one photon in the
time interval~$\tG$ (Fig.~\ref{fig:Wed24May120508BST2023}(b)). Another
equivalent way, thanks to the properties of Poisson processes, is to
add a time gap~$\tG$ after each photon
(Fig.~\ref{fig:Wed24May120508BST2023}(c)).  Poisson emission being
independent at all times of what happened previously, the addition
``by-hand'' of a gap does not alter the probability of emission after
the gap. In all cases, the sources should be normalized to their
emission rate which, if the original, underlying Poisson process has
emission rate~$\gamma$, is then, by opening a gap of duration~$\tG$,
given by:
\begin{equation}
  \label{eq:Wed24May120830BST2023}
  \gamma_\mathrm{G}={\gamma\over1+\gamma\tG}\,.
\end{equation}
There are various ways to obtain this result. From the numerically
more efficient scheme in Fig.~\ref{fig:Wed24May120508BST2023}(c)), one
sees that for a total number of events (``clicks'')~$N_c$, emitted at
a rate~$\gamma$ so that the last photon is detected at
time~$t_\mathrm{last}=N_c/\gamma$ (first at~$t_\mathrm{first}=0$),
adding a time gap subsequent to each photon leaves us with an emission
rate $\gamma_\mathrm{G}=N_c/(t_\mathrm{last}+N_c\tG)$ which is,
dividing numerator and denominator by~$t_\mathrm{last}$,
Eq.~(\ref{eq:Wed24May120830BST2023}). Clearly, opening a gap lowers
the intensity, a point to which we shall return later on.

Such a gap, for which $g^{(2)}(\tau)=0$ not only at~$\tau=0$ but for
some finite time window, clearly solves the issue of finite-time
resolution: as long as our detector is more accurate in time
than~$\tG$, it will never accidentally collect two or more photons at
once. A gap is precisely what makes a superconductor a perfect
conductor~\cite{lopezcarreno22a}, although measurements also come with
imperfections for an electric current as well. One can still cancel
the resistivity exactly (meaning that deviations from zero cannot be
measured). Photons being precisely quanta, so discrete quantities, it
should be expected that one could similarly achieve $g^{(2)}(0)=0$
exactly for light, meaning that one could generate a perfect stream of
single-photons without ever observing more than one at a time for as
long as we count.  So far, instead, one could ``only'' considerably
suppress, but not eliminate completely,
coincidences~\cite{schweickert18a,hanschke18a} (this can still be
claimed mathematically~\cite{huber20a}).

\begin{figure}
  \includegraphics[width=.9\linewidth]{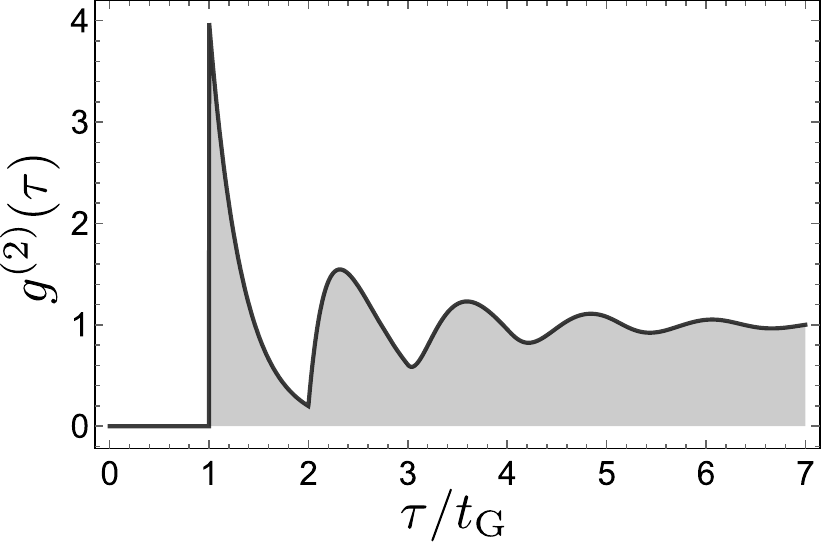}
  \caption{Glauber's second-order correlation function $g^{(2)}(\tau)$
    for a gapped single-photon source with~$\gamma\tG=3$ (shown for
    positive~$\tau$ only, the function is symmetric). The
    single-photon gap is followed by strong correlations of the
    bunching type which are damped in time.}
  \label{fig:Fri26May200445BST2023}
\end{figure}

An interesting property of such a gapped stream of photons is that,
while it satisfies Eq.~(\ref{eq:Sat22Apr190318BST2023}) by
construction, later photons do not ``remain'' uncorrelated with the
one that opens the gap, i.e., we do not have
Eq.~(\ref{eq:Sat22Apr190411BST2023}) as one could expect but, instead:
% %
\begin{widetext}
  \begin{equation}
    \label{eq:Wed7Jun143947BST2023}
    g^{(2)}(\tau) = (1+\gamma \tG)e^{-\gamma(|\tau|-\tG)}\sum^{\infty}_{n = 0} \frac{\big[\gamma\big(|\tau| - (n+1)\tG\big)e^{\gamma \tG}\big] ^{n}}{n!}\mathbb{1}_{[(n+1)\tG, \infty [} (|\tau|)
  \end{equation}
\end{widetext}
where $\mathbb{1}_A(t)$ is the indicator function, i.e., it is~1 if
$t\in A$ and~0 otherwise.  An example of such correlations is shown in
Fig.~\ref{fig:Fri26May200445BST2023}. There is a superbunching peak
immediately after the gap and then damped oscillations with a
frequency approximately given by Eq.~(\ref{eq:Wed24May120830BST2023}).
This is an interesting feature given that $g^{(2)}(\tau)$ measures a
density of two-photon events, regardless of whether photons are
successive ones or if they are others in between them: it correlates
each photon to every other photon. Therefore, that a photon
at~$\tau=\tG$ ``becomes'' correlated with a photon at~$\tau=0$ by the
mere fact that those in between have been removed could be
surprising. Quantum opticians know well that~$g^{(2)}(\tau)$ is
insensible to losses and whether all photons are detected or only a
tiny amount of them does not change their measurement (only the time
they need to acquire it with the noise-to-signal ratio they want).
Another way to describe this apparent paradox is that from an
initially uncorrelated stream of photons, one can produce positive
correlations of a bunching type by \emph{removing} photons, and adding
nothing.

The reason for these strong positive correlations can be linked to the
Monty Hall problem~\cite{selvin75b} that is a famous paradox whereby
providing apparent useless information to a decision-problem turns out
to alter its chance of success, when someone has knowledge of how to
provide uselessly this information. Namely, in the Monty Hall problem,
a player is given a choice between three doors, one of which conceals
a prize. After selecting a door, someone with knowledge of which door
hides the prize opens one of the two unselected doors (without
prize). Should the player stick to their initial choice or switch to
the remaining door to win the prize? Counter-intuitively, the player,
who had initially a probability 1/3 to find the prize, now has
probability 2/3 if they switch the door. This is surprising at first
since, whatever the initial choice, it is always possible to open a
loosing door, so no information seems to have been revealed.

In our case, enforcing the single-photon attribute in a given
time-window would similarly seem to be unrelated to photon
correlations elsewhere. Why would a photon from an initially
uncorrelated stream become more likely to be detected immediately at
the time at which we have been revealead that no other photons are
present?  It would indeed be the case that no such dramatic
correlations occur if the subtracted photons would have been so by
chance, and not deterministically (just like the Monty hall paradox
breaks if the door is opened randomly).  Then~$g^{(2)}(\tau)$ would
remain identity everywhere. We find, instead, that enforcing
Eq.~(\ref{eq:Sat22Apr190318BST2023}) comes at the price of propagating
correlations at all times according to
Eq.~(\ref{eq:Wed7Jun143947BST2023}).

We now proceed to explain the origin for this result which, like the
Monty Hall problem, although maybe counter-intuitive at first, is
actually easily and unequivocally understood by working out the
corresponding probabilities (in the Monty Hall case it is interesting
that it generated considerable opposition, in particular as its
popularization came from a woman~\cite{vossavant90a}, and that even
Erd\"os himself was not accepting the result~\cite{vazsonyi99a}). In
our case, one can observe that the stronger correlations are in the
second time-window following the gap one, i.e.,
for~${\tG\le t\le 2\tG}$, where one has, furthermore, a simple
exponential decay. This is because in this window, we have one and
only one photon (if there is no photon, then this does not contribute
to signal as already pointed out, and there can be no two photons in a
time window of width~$\tG$ by construction), that remains Poisson
distributed since any possibly-removed photon was in the previous time
window $t<\tG$ and the one left felt no influence from the memoryless
nature of Poisson processes.  We have therefore, in this window, the
$g^{(2)}(\tau)$ of an heralded single photon (the ``heralder'' being
the first detected photon which opens the gap window) which, if its
own distribution of time emission is a simple exponential process, as
is the case for random emission, also decays
exponentially~\cite{zubizarretacasalengua23a}. It is also important
that no other photon be present in the second time window, as one
would have otherwise an other scenario of an heralded $N$-photon
emission~\cite{khalid_arXiv23a} which is multi-exponential.

To work out the correlations at subsequent times, we follow the route
opened by Reynaud~\cite{reynaud83a} of working out photon correlations
from conditional probabilities of the emission processes, and by Kim
\emph{et al.}~\cite{kim87a} who further established the relationship
between the waiting time distribution~$w(\tau)$ between
\emph{successive} photons and Glauber's $g^{(2)}(\tau)$ between any
two photons:
\begin{equation}
  \label{eq:Wed7Jun105220BST2023}
   \gamma_\mathrm{G} g^{(2)}(\tau)=w(\tau)+(w* w)(\tau)+(w*w*w)(\tau)+\cdots
\end{equation}
where~$*$ denotes convolution, i.e., for the second term of this
infinite series:
\begin{equation}
  \label{eq:Wed7Jun105312BST2023}
  (w*w)(\tau)=\int_0^\infty w(\tau')w(\tau-\tau')\,d\tau'
\end{equation}
where the upper limit for integration could have been stopped
at~$\tau$ since~$w$ is zero for negative arguments, in which case this
provides the probability of detecting a photon at~$t=0$ and another
one at $t=\tau$ with one and exactly one photon in between at time
$t=\tau'$, averaged over all possible values of $\tau'$, i.e., this
provides the waiting-time distribution~$w_2$ for the second
photon.

In our case, the waiting time distribution is a good starting point
since this is simply for our gapped coherent source:
\begin{equation}
  \label{eq:Wed24May153030BST2023}
    w(\tau)=\gamma e^{-\gamma(\tau-\tG)}\mathbb{1}_{[\tG,\infty[}(\tau)\,.  
\end{equation}
This follows again from the memoryless properties of Poisson point
processes, and is an obvious consequence of the equivalent
construction as sketched in Fig.~\ref{fig:Wed24May120508BST2023}(c)
(the sum of two Poisson processes is also a Poisson process).  Note
that unlike $g^{(2)}(\tau)$ that measures the density of any
two-photon events, the waiting-time distribution that quantifies
successive events is actually intuitive.

The waiting-time distributions are thus obtained by computing
Eq.~(\ref{eq:Wed7Jun105312BST2023}) for
Eq.~(\ref{eq:Wed24May153030BST2023}), i.e.,
\begin{multline}
  \label{eq:Wed7Jun190725BST2023}
  w_2(\tau)=\gamma^2\int_{0}^{\tau} e^{-\gamma(\tau'-\tG)}e^{-\gamma(\tau-\tau'-\tG)}\times\\\times\mathbb{1}_{[\tG,\infty[}(\tau')\mathbb{1}_{[\tG,\infty[}(\tau-\tau')\,d\tau'\,.
\end{multline}
This would be a simple convolution of exponentials and thus a trivial
result, except that the indicator functions impose various
restrictions on the time domains, namely, they require $\tau$ to be
larger than~$2\tG$ and $\tau'$ to remain between~$\tG$ and~$\tau-\tG$:
\begin{equation}
  \label{eq:Wed7Jun153812BST2023}
  w_2(\tau\ge 2\tG)=\gamma^2\int_{\tG}^{\tau-\tG} e^{-\gamma(\tau'-\tG)}e^{-\gamma(\tau-\tau'-\tG)}\,d\tau'
\end{equation}
so that finally, one has:
\begin{equation}
  \label{eq:Wed7Jun191954BST2023}
  w_2(\tau)=\gamma^2(\tau-2\tG)e^{-\gamma(\tau-2\tG)}\mathbb{1}_{[2\tG,\infty[}(\tau)\,.
\end{equation}
Subsequent cases are similarly obtained from the higher-order
convolutions, corresponding to the successive single photons isolated
in their respective gap windows. The general case is obtained as the
$n$th convolution which can be calculated by recurrence:
\begin{equation}
  \label{eq:Wed7Jun192640BST2023}
  w_n(\tau)=w^{*n}(\tau)=(w^{*(n-1)}*w)(\tau)
\end{equation}
with similar constrains on the domains of integrations, now involving
the product of indicator functions
$\mathbb{1}_{[n\tG,\infty[}(\tau)\mathbb{1}_{[\tG,\infty[}(\tau-\tau')$
to yield
\begin{equation}
  \label{eq:Wed7Jun192918BST2023}
  w_n(\tau)=\gamma\int_{(n-1)\tG}^{\tau-\tG}w_{n-1}(\tau')e^{-\gamma(\tau-\tau'-\tG)}\,d\tau'
\end{equation}
which provides the $n$th photon waiting time distribution
\begin{equation}
  \label{eq:Thu8Jun100359BST2023}
  w_n(\tau)=\frac{\gamma^n\big(\tau - n\tG\big)^{n-1}e^{-\gamma(\tau-n \tG)}}{(n-1)!}\mathbb{1}_{[n\tG, \infty [} (\tau)      
\end{equation}
from which our main result Eq.~(\ref{eq:Wed7Jun143947BST2023}) is
obtained through Eq.~(\ref{eq:Wed7Jun105220BST2023}) (including for
$t\le\tG$) as
$g^{(2)}(\tau)=\gamma_\mathrm{G}^{-1}\sum_{n=1}^\infty
w_n(|\tau|)$. Note that the actual emission rate of the stream, so
$\gamma_\mathrm{G}$, provides the link between~$g^{(2)}$ and~$w$ in
Eq.~(\ref{eq:Wed7Jun105220BST2023}) while the original rate~$\gamma$
from the underlying Poisson process enters in the waiting time
distribution~(\ref{eq:Wed24May153030BST2023}), so that the maximum amount of
correlations, at the time gap is given by
\begin{equation}
  \label{eq:Thu8Jun105219BST2023}
  g^{(2)}(\tG)={\gamma\over\gamma_\mathrm{G}}=1+\gamma\tG\,.
\end{equation}
This is also obtained directly from
Eq.~(\ref{eq:Wed7Jun143947BST2023})
as~$\lim_{\tau\to\tG\atop\tau>\tG}g^{(2)}(\tau)$.  Note also that
$g^{(2)}(\tG)$ is not actually defined, or presents a discontinuity
there, since $\lim_{\tau\to\tG\atop\tau<\tG}g^{(2)}(\tau)=0$ and
Eq.~(\ref{eq:Thu8Jun105219BST2023}) really refers to the
% ${\tau\to\tG\atop\tau>\tG}$
$\tau>\tG$ limit. The correlations are maximized by large values of
$\gamma\tG$, which is also the mean number of photons in the time gap
(before removing the extra ones in
Fig.~\ref{fig:Wed24May120508BST2023}(a)). This brings us back to the
Monty Hall problem where the amount of deterministically removed
information is linked to the magnitude of the effect (an insightful
argument from vos Savant was to generalize the game to a large number
of doors~\cite{vossavant90a} making the probability to win by
switching converge to one). 

From these results, one can now work out the conditions to produce a
SPS that satisfies Eqs.~(\ref{eq:Sat22Apr190318BST2023})
and~(\ref{eq:Sat22Apr190411BST2023}): this is obtained in the limit of
no signal, since from Eq.~(\ref{eq:Wed24May120830BST2023}) one must
have $\gamma\tG=0$ or, for sizable~$\tG$, a vanishing~$\gamma$.  A
perfect SPS with no extra-correlations is thus one with very scarce
signal: a dim laser (coherent or Poissonian) from which Poisson bursts
are brushed off with the gap, that, in most cases, will do nothing as
there are so few clicks anyway, but whenever two (or more) clicks
would get close by chance, the extra ones are removed. In fact,
Fig.~\ref{fig:Wed24May130615BST2023}(b) has been obtained in this way
with~$\gamma\tG=10^{-3}$ and is thus a physical~$g^{(2)}$ (the
departure to the ideal case is not visible to the eye).

\section{Pulsed single-photon source}
\label{sec:Sun23Apr104328BST2023}

Another limit to contrast to $\gamma\tG\to 0$ in
Fig.~\ref{fig:Wed24May130615BST2023}(b), is that where
$\gamma\tG\gg 1$, in which case, although the signal is still,
eventually, stationary, its correlations are so strong as to be
reminiscent of the opposite regime of single-photon emission: under
pulsed excitation.

The pulsed SPS consists in generating each photon individually with
pulses of excitation that drive the system in its excited
state~\cite{brunel99a,michler00a}. The ideal limit consists of a train
of equally spaced photons with a Dirac comb of two-photon
correlations:
\begin{equation}
  \label{eq:Sun23Apr105018BST2023}
  g^{(2)}(\tau)=\tG\sum_{n\neq 0}\delta(\tau-n\tG)
\end{equation}
where the time gap is now externally imprinted by the laser pulses.
The peak averages are normalized:
\begin{equation}
  \label{eq:Thu8Jun143701BST2023}
  {1\over\tG}\int_\mathrm{peak} g^{(2)}(\tau)\,d\tau=1\,.
\end{equation}
Considering fluctuations, or \emph{jitter}, of each photon emission
around their reference~$n\tG$ emission time, according to a
distribution~$D(t)$, we find:
\begin{equation}
  \label{eq:Thu8Jun141902BST2023}
  g^{(2)}(\tau)=\tG\sum_{n\neq 0}(D \star D)(|\tau-n\tG|)
\end{equation}
where $\star$ represents this time the auto-correlation:
\begin{equation}
  \label{eq:Wed21Jun123013BST2023}
  (D\star D)(\tau)\equiv\int_{0}^{+\infty} D(t)D(t+\tau)\,d t\,.
\end{equation}

\begin{figure}
  \includegraphics[width=.9\linewidth]{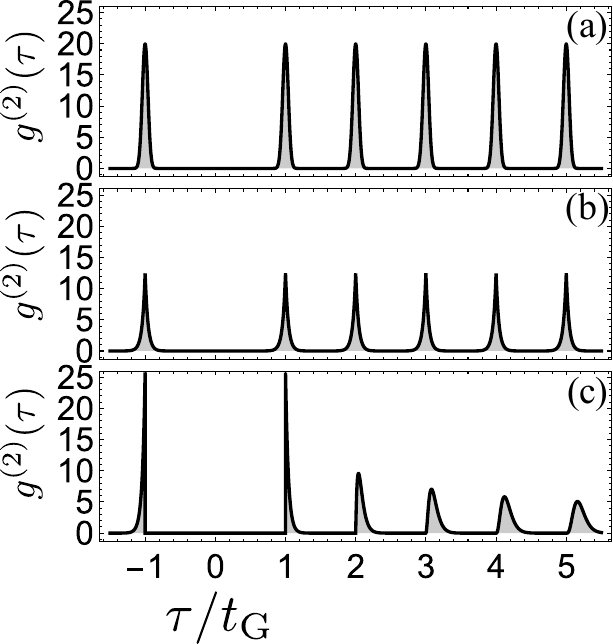}
  \caption{Two-photon correlations in (a, b) the pulsed regime and (c)
    the stationary regime for a gapped SPS with~$\gamma\tG=25$. The
    pulses would eventually overlap, wash out and plateau to~1 in the
    latter case. In the pulsed case, the suppressed peak at~$\tau=0$
    is the only signature of two-photon suppression while the full
    structure is a testimony to that in~(c).}
  \label{fig:Thu8Jun142758BST2023}
\end{figure}

We show, in Fig.~\ref{fig:Thu8Jun142758BST2023} the cases of (a)
normally-distributed fluctuations, i.e.,
with~$D(t)={\sqrt{{2\gamma^2}/{\pi}}{e^{-\frac{1}{2}(\gamma t)^2}}}$,
and (b) of asymmetric exponentiel decay with
$D(t)={\gamma e^{-\gamma t}}$ along with, in panel~(c),
Eq.~(\ref{eq:Wed7Jun143947BST2023}) for~$\gamma\tG\gg 1$.  The
resemblances are clear as well as the reason why: in such a case, the
background Poisson signal is so strong that its gapping is, indeed,
tentamount to creating a periodic array of single-photons since for a
sufficiently bright signal, there is always a random photon close
enough to the start of the next gapping window, after all those extra
from the previous window have been trimmed-off.  Clearly this produces
a stream similar to a sequence of equally-spaced pulses.  This is in
fact another way to dissipate immediately the counter-intuitive
features of the gapped stream: seeing it as a stationary version of
the pulsed SPS, where the ordering is not externally imprinted, but
internally by the gapping of the photons. It must have a finite
coherence-time since, as for all stationary signals, $g^{(2)}(\tau)=1$
for~$\tau$ large enough and so there must be a dissolving of the peaks
into a flat backround. The reason for this return to stationarity is
due to the accumulated effect of time fluctuations for each peak in
the gapped case while, when they come from external pulses, they are
reset for each peak. This results in several differences, even in the
strongly correlated region.  Foremost, since the waiting time
distribution Eq.~(\ref{eq:Thu8Jun100359BST2023}) is normalized, in the
limit where the peaks are greatly separated, they also fulfill
Eq.~(\ref{eq:Thu8Jun143701BST2023}). Figure~\ref{fig:Thu8Jun142758BST2023}(c)
in fact shows the~$w_n(\tau)$ in isolation.  In stark contrast with
the pulse case, however, there is a decay of the maxima of the peaks
with an envelope for the comb. Such a decay can also be found in some
pulsed-excitation schemes, but for a different reason. Santori
\emph{et al.}~\cite{santori01a} indeed observed (experimentally) a
decaying envelope for the peaks, which they identified was caused by
blinking, i.e., intermittency of the emission, a well-known problem of
solid-state emitters. This furthermore concerns, precisely, the
\emph{areas} of the peaks, which decay, while in our case, the
\emph{maximum} of the peaks, that otherwise remain normalized, decay,
until the peaks broaden to a point that they become unresolvable and a
flat~$g^{(2)}(\tau)=1$ is recovered (stationarity). Another striking
difference is that pulsed autocorrelations feature symmetric peaks,
since whatever shape the jitter distribution~$D$ takes, its
auto-correlation~Eq.~(\ref{eq:Wed21Jun123013BST2023}) symmetrizes the
photon coincidence, that can be equally delayed or postponed. This is
seen in Figs.~\ref{fig:Thu8Jun142758BST2023}(a) and~(b) that stem from
both symmetric and asymmetric fluctuations.  In contrast, the gapped
SPS is markedly asymmetric, with photons being delayed only, since the
gap is, by construction, impenetrable. As another consequence, the
frequency increasingly lags in time.

\section{Discussion}

Our starting point was the requirement of a stream of photons that
ensures that never two photons be found together. Let alone other
important attributes for many applications (such as
indistinguishability or
brightness~\cite{ding16a,somaschi16a,aharonovich16a} that are
crucial for applications~\cite{wang19c}), this is a universal and
basic requisite for a single-photon source.  We have discussed how
this becomes a fundamental problem for the paradigmatic cases of a
two-level system spontaneously emitting, when including their
detection---which is central in a quantum theory---as this two-photon
suppression reduces to one-time only ($\tau=0$). Demanding such a
suppression over a finite interval of time---a time gap---we found
interesting properties of the resulting signal that shed light on the
meaning and nature of single-photon emission. Foremost, we find
temporal two-photon correlations, which we have described exactly. The
phenomenology is interesting because such oscillations are commonplace
with strongly-driven two-level systems, and are usually associated to
Rabi dynamics. We find in contrast (not opposition), that regardless
of the mechanism to produce photons, the mere presence of the gap
condition~Eq.~(\ref{eq:Sat22Apr190318BST2023}) results in such
oscillations as the emergence of an ordering of the photons in a
sequence. This is a property to keep in mind when considering highly
antibunched sources: such oscillations are the manifestation of such
good single-photon emission that even in a stationary regime, it
acquires some properties of the pulsed regime. This further shows that
brightness, i.e., the intensity of the emission, that is not or not
well captured by~$g^{(2)}(\tau)$, is an important aspect of
single-photon emission, and not only for considerations of signal but
as a result of photon ordering. In contrast, indistinguishability
appears to remain less connected to the problem of suppression of
multiple photons, although in the case of an incoherent two-level
system, we have shown that one can realize an ideal perfect SPS at the
cost of a complete loss of indistinguishability. We also emphasized
the conter-intuitive nature of $g^{(2)}(\tau)$ as opposed to other,
but less popular, quantities such as the waiting-time
distribution. Although the peculiarities we have focused on are
quickly understood, it is easy to cook-up further counter-intuitive
features. As another illustration of this, if in our gaping procedure
one does not always remove the extra photons but does so with
probability~$p$ only, instead of the perfect
antibunching~(\ref{eq:Sat22Apr190318BST2023}), one now produces actual
coincidences for~$|\tau|<\tG$ from the underlying uncorrelated stream,
namely, $g^{(2)}(\tau)=(1-p)(p \gamma \tG +1)e^{- p \gamma|\tau|}$
which thus turns to bunching~$g^{(2)}(0)>1$
for~$p<1-\frac{1}{\gamma \tG}$. This is always possible (positive~$p$)
as long as $\gamma\tG\ge 1$, i.e., there should be more than one
photon on average in the time gap. In this case, the maximum bunching
is $(1 + \gamma\tG)^2/(4\gamma\tG)$ for the
probability~$p={1\over2}(1-{1\over\gamma\tG})$ to remove photons
closer than~$\tG$.  This tends to infinite bunching for 50\% removal
of the photons in the gap, in the limit of large~$\gamma\tG$. Again,
how removal of photons---this time probabilistic as opposed to
systematic---from an uncorrelated stream, can result in arbitrarily
large and genuine bunching (as opposed to antibunching), may be
surprising at first.  The reader should however be able to explain it
and provide the full~$g^{(2)}(\tau)$ for this case as well following
our above discussion~\cite{khalid_thesis24a}.

Our approach also provides another and insightful picture of the
standard SPS as realized by an incoherently-pumped two-level system:
this can be understood similarly to the scheme of
Fig.~\ref{fig:Wed24May120508BST2023}(c) but instead of ``inserting'' a
constant time gap~$\tG$ after each photon from the underlying
uncorrelated stream, one instead inserts a Poisson-distributed delay
(i.e., following the exponenial distribution with
parameter~$\gamma'$). Then one obtains the two-photon correlation
function of the incoherent single-photon emitter
$g^{(2)}_\mathrm{2LS}(\tau)=1-\exp\big(-(\gamma+\gamma')|\tau|\big)$
(we remind that $\gamma$ is the emission rate of the uncorrelated
backbone emission) for a total emission rate of the SPS thus formed
of~$\gamma\gamma'/(\gamma+\gamma')$ according to the arguments given
to obtain Eq.~(\ref{eq:Wed24May120830BST2023}). This provides a
numerically easy and efficient way to generate such a
stream~\cite{lopezcarreno18a} obtained under either the condition
$P_\sigma=\gamma$ and~$\gamma_\sigma=\gamma'$ or the symmeric
$P_\sigma=\gamma'$ and~$\gamma_\sigma=\gamma$.  Such a picture should
also make even more clear why an incoherently-excited two-level system
fails to provide perfect single-photons: its suppression of
coincidences stems from the feeblest type of repulsion between
photons, with indeed the highest probabilities for smaller delays and
whose underlying emission is furthemore prone to Poisson bursts, as
shown in Fig.~\ref{fig:Wed24May120508BST2023}(a). This is the reason
for the steep $g^{(2)}(\tau)$ and why its $\tau=0$ coincidences can
never vanish entirely at a fundamental level: initially closely-spaced
photons can be further unlucky to not be much separated by their
Poisson repulsion. While this combination of events is indeed
unlikely, it is doomed to happen over a sufficient large sampling (one
instance is captured from actual simulation for the seventh and eighth
photons in streams~(a) and~(d) of
Fig.~\ref{fig:Wed24May120508BST2023}).

Our discussion also tampers down the statement that it ``is not enough
to have single-photon wavepackets to have single-photon light'' but
that one ``need[s] in addition to know at which time each
single-photon wavepacket is emitted''~\cite{aspect13b}. We have just
exhibited a source from which one has no long-term knowledge of when
each photon is to be emitted, but that, locally, is even more
efficient than existing sources in suppressing coincidences.

The problem of how to actually engineer a quantum emitter that would
produce such a gapped coherent state is another question. We want to
highlight, however, that the joint subnatural linewidth antibunched
sourced proposed in Ref.~\cite{lopezcarreno18b} bears some resemblance
with our idealized case and was indeed the motivation for this
work. Namely, there is a plateaued $g^{(2)}(\tau)$ which, although not
strictly zero, is still much flatter than the incoherently-driven
two-level system. Furthermore, its emission is obtained at the price
of a considerable drop in the signal, indeed removing---not extra
photons in a time gap---but the coherent fraction of the signal, again
bringing the question of brightness into the shape of the
correlations. Contrasting this case---which is, this time, the result
of an explicit dynamical model---with our previous discussion,
highlights again how the imperfect single-photon character of the
incoherently pumped two-level system stems from the feeble repulsion
between photons.  In contrast, a distribution of time delays that is
minimum at~$\tau=0$ (instead of maximum like for the incoherent SPS),
e.g., of the Maxwell distribution type (from the kinetic theory of
gases), does produce a plateaued $g^{(2)}(\tau)$ much like the one
found in the coherently homodyned emitter~\cite{lopezcarreno18b}. This
raises further questions on the nature and origin of antibunching in
resonance fluorescence~\cite{hanschke20a} and indeed invites to design
schemes to more strongly repel
photons---qualitatively~\cite{zubizarretacasalengua20a,meyerscott20a}
in the way each photon pushes away the next ones as described
by~$g^{(2)}(\tau)$---instead of ``merely''
quantitatively~\cite{reimer19a,thomas21a}---reducing~$g^{(2)}(0)$---so
as to eventually implement a perfect single-photon source.

Finally, our discussion of temporal correlations of photons relates to
a popular problem in condensed matter physics, namely, of the
so-called pair correlation function $g(r)$ for the distance~$r$ (in
space), also known as the radial distribution function, of
fluids~\cite{kirkwood39a}. This has long been known from the
diffraction of X-rays to exhibit damped oscillations, with shapes that
are reminiscent of those we have described above. Our result
Eq.~(\ref{eq:Wed7Jun143947BST2023}) is indeed found in this context in
the extreme idealization for a liquid as noninteracting hard-spheres
in one dimension (so ``hard rods''). This has been derived by several
authors using a variety of techniques, from probabilistic arguments of
the Clausius or Smoluchowski type for the distribution of free paths
of a molecule in a gas~\cite{zernike27a,frenkel_book46a}, to integral
equations derived from the partition
function~\cite{kirkwood39a,salsburg53a,sells53a,wertheim63a}, where
equations~(\ref{eq:Sat22Apr190318BST2023}--\ref{eq:Sat22Apr190411BST2023})
are known as the ``hole-correction approximation''.  Our own solution
is obtained from the probabilistic side, but invoking arguments of
spontaneous emission from heralded photons instead of configurations
of points distributed spatially on a finite line before taking the
thermodynamic limit. Such direct eventual relationships bring the
question whether thermodynamic concepts as well as results and
techniques of statistical physics could shed light on photon
correlations of increasingly complex types of quantum light, and if
one should characterize photon streams as temporal Tonks gases, with
an equation of state to relate their entropy with other macroscopic
variables such as their pressure and compressibility, albeit, again,
in time as opposed to in space.  With this parallel, one could thus
understand the case of Fig.~\ref{fig:Thu8Jun142758BST2023}(c) as a
photon (temporal) crystal and that of
Fig.~\ref{fig:Fri26May200445BST2023} as its melting into a photon
(temporal) fluid. Whether such parallels can benefit one or the other
discipline remains to be ascertained but their conceptual proximity
cannot be denied, and old results obtained under drastic
approximations for fluids may well be resurrected as fundamental for
photons which are intrinsically non-interacting and, temporally,
one-dimensional.

\section{Summary \& Conclusions}
\label{sec:Sun23Apr110641BST2023}

We have taken a point-process approach to photon correlations, by
focusing on the signal as ``clicks'' recorded in time. This has
allowed us to formulate of concept of a perfect SPS in the form of a
gapped coherent state, which is a Poisson (stationary) stream in which
a gapping condition is enforced that no two photons are ever found
closer than the time gap~$\tG$.  This ensure that $g^{(2)}(0)$ can
actually be measured as exactly zero even with imperfect detectors or
taking into account fundamental time uncertainties from the
measurement, and thus also enforce the vanishing of coincidences of
higher-numbers of photons (that are, counter-intuitively, possible to
any degree if $g^{(2)}(0)$ is merely very small but nonzero).  We have
shown that this results in strong correlations in the form of
oscillations that alternate bunching and antibunching, with, in
particular, increased probabilities to find a photon immediately after
the gap, followed by damped oscillations. We have explained the nature
of the phenomenon and provided closed-form analytical expressions for
it. We also identified various fundamental attributes in the nature of
single-photon emission, including the constraints imposed by the
signal (intensity) with the result of tranposing our stationary source
into one with attributes of pulsed emission, thereby establishing a
fundamental connection between continuous-wave and pulsed
single-photon sources.  This allowed us to revisit the nature of the
paradigmatic single-photon emission from the incoherent emission of a
two-level system, that similarly introduces, not a rigid gap, but an
exponentially distributed delay between photons.  The relationship of
the gapped coherent state to the recently proposed scheme of joint
antibunched and subnatural-linewidth emission, and still other
emitters able to realize other types of photon repulsions, as well as
other phases of matter, could be fruitfully pursued.

\begin{acknowledgments}
  We thank Ceri Neale and Luke Collins, who first observed the
  phenomenon as part of their Research module at the University of
  Wolverhampton, as well as Dylan Weston, Joel Warner, Damien Pitts,
  Luke Bickford, Elena del Valle, Eduardo Zubizarreta Casalengua and
  Camilo L\'opez Carre\~no for comments and discussions.  We thank in
  particular Elena for bringing our attention to the pair correlation
  function of hard rods.
\end{acknowledgments}

\bibliographystyle{naturemag}
\bibliography{sci,Books,arXiv} 

\begin{thebibliography}{10}
\expandafter\ifx\csname url\endcsname\relax
  \def\url#1{\texttt{#1}}\fi
\expandafter\ifx\csname urlprefix\endcsname\relax\def\urlprefix{URL }\fi
\providecommand{\bibinfo}[2]{#2}
\providecommand{\eprint}[2][]{\url{#2}}

\bibitem{obrien09a}
\bibinfo{author}{O'Brien, J.~L.}, \bibinfo{author}{Furusawa, A.} \&
  \bibinfo{author}{\Vuckovic, J.}
\newblock \bibinfo{title}{Photonic quantum technologies}.
\newblock \emph{\bibinfo{journal}{Nature Phys.}} \textbf{\bibinfo{volume}{3}},
  \bibinfo{pages}{687} (\bibinfo{year}{2009}).
\newblock \urlprefix\url{doi:10.1038/nphoton.2009.229}.

\bibitem{senellart17a}
\bibinfo{author}{Senellart, P.}, \bibinfo{author}{Solomon, G.} \&
  \bibinfo{author}{White, A.}
\newblock \bibinfo{title}{High-performance semiconductor quantum-dot
  single-photon sources}.
\newblock \emph{\bibinfo{journal}{Nature Nanotech.}}
  \textbf{\bibinfo{volume}{12}}, \bibinfo{pages}{1026} (\bibinfo{year}{2017}).
\newblock \urlprefix\url{doi:10.1038/nnano.2017.218}.

\bibitem{couteau23a}
\bibinfo{author}{Couteau, C.} \emph{et~al.}
\newblock \bibinfo{title}{Applications of single photons to quantum
  communication and computing}.
\newblock \emph{\bibinfo{journal}{Nature Rev. Phys.}}
  \textbf{\bibinfo{volume}{5}}, \bibinfo{pages}{326} (\bibinfo{year}{2023}).
\newblock \urlprefix\url{doi:10.1038/s42254-023-00583-2}.

\bibitem{couteau23b}
\bibinfo{author}{Couteau, C.} \emph{et~al.}
\newblock \bibinfo{title}{Applications of single photons in quantum metrology,
  biology and the foundations of quantum physics}.
\newblock \emph{\bibinfo{journal}{Nature Rev. Phys.}}
  \textbf{\bibinfo{volume}{5}}, \bibinfo{pages}{354} (\bibinfo{year}{2023}).
\newblock \urlprefix\url{doi:10.1038/s42254-023-00589-w}.

\bibitem{hanburybrown56c}
\bibinfo{author}{{Hanbury Brown}, R.} \& \bibinfo{author}{Twiss, R.~Q.}
\newblock \bibinfo{title}{The question of correlation between photons in
  coherent light rays}.
\newblock \emph{\bibinfo{journal}{Nature}} \textbf{\bibinfo{volume}{178}},
  \bibinfo{pages}{1447} (\bibinfo{year}{1956}).
\newblock \urlprefix\url{doi:10.1038/1781447a0}.

\bibitem{glauber63a}
\bibinfo{author}{Glauber, R.~J.}
\newblock \bibinfo{title}{Photon correlations}.
\newblock \emph{\bibinfo{journal}{Phys. Rev. Lett.}}
  \textbf{\bibinfo{volume}{10}}, \bibinfo{pages}{84} (\bibinfo{year}{1963}).
\newblock \urlprefix\url{doi:10.1103/PhysRevLett.10.84}.

\bibitem{zubizarretacasalengua17a}
\bibinfo{author}{{Zubizarreta Casalengua}, E.}, \bibinfo{author}{{L\'opez
  Carre\~no}, J.~C.}, \bibinfo{author}{del Valle, E.} \&
  \bibinfo{author}{Laussy, F.~P.}
\newblock \bibinfo{title}{Structure of the harmonic oscillator in the space of
  $n$-particle {Glauber} correlators}.
\newblock \emph{\bibinfo{journal}{J. Math. Phys.}}
  \textbf{\bibinfo{volume}{58}}, \bibinfo{pages}{062109}
  (\bibinfo{year}{2017}).
\newblock \urlprefix\url{doi:10.1063/1.4987023}.

\bibitem{grunwald19a}
\bibinfo{author}{Gr\"unwald, P.}
\newblock \bibinfo{title}{Effective second-order correlation function and
  single-photon detection}.
\newblock \emph{\bibinfo{journal}{New J. Phys.}} \textbf{\bibinfo{volume}{21}},
  \bibinfo{pages}{093003} (\bibinfo{year}{2019}).
\newblock \urlprefix\url{doi:10.1088/1367-2630/ab3ae0}.

\bibitem{arXiv_lopezcarreno16c}
\bibinfo{author}{Carre{\~n}o, J. C.~L.}, \bibinfo{author}{Casalengua, E.~Z.},
  \bibinfo{author}{del Valle, E.} \& \bibinfo{author}{Laussy, F.~P.}
\newblock \bibinfo{title}{Criterion for single photon sources}.
\newblock \emph{\bibinfo{journal}{arXiv:1610.06126}}  (\bibinfo{year}{2016}).

\bibitem{molmer93a}
\bibinfo{author}{M{\o}lmer, K.}, \bibinfo{author}{Castin, Y.} \&
  \bibinfo{author}{Dalibard, J.}
\newblock \bibinfo{title}{{Monte Carlo} wave-function method in quantum
  optics}.
\newblock \emph{\bibinfo{journal}{J. Opt. Soc. Am. B}}
  \textbf{\bibinfo{volume}{10}}, \bibinfo{pages}{524} (\bibinfo{year}{1993}).
\newblock \urlprefix\url{doi:10.1364/JOSAB.10.000524}.

\bibitem{delvalle12a}
\bibinfo{author}{del Valle, E.}, \bibinfo{author}{Gonz\'alez-Tudela, A.},
  \bibinfo{author}{Laussy, F.~P.}, \bibinfo{author}{Tejedor, C.} \&
  \bibinfo{author}{Hartmann, M.~J.}
\newblock \bibinfo{title}{Theory of frequency-filtered and time-resolved
  $n$-photon correlations}.
\newblock \emph{\bibinfo{journal}{Phys. Rev. Lett.}}
  \textbf{\bibinfo{volume}{109}}, \bibinfo{pages}{183601}
  (\bibinfo{year}{2012}).
\newblock \urlprefix\url{doi:10.1103/PhysRevLett.109.183601}.

\bibitem{lopezcarreno22a}
\bibinfo{author}{{L\'opez Carre\~no}, J.~C.}, \bibinfo{author}{Casalengua,
  E.~Z.}, \bibinfo{author}{Silva, B.}, \bibinfo{author}{del Valle, E.} \&
  \bibinfo{author}{Laussy, F.~P.}
\newblock \bibinfo{title}{Loss of antibunching}.
\newblock \emph{\bibinfo{journal}{Phys. Rev. A}}
  \textbf{\bibinfo{volume}{105}}, \bibinfo{pages}{023724}
  (\bibinfo{year}{2022}).
\newblock \urlprefix\url{doi:10.1103/PhysRevA.105.023724}.

\bibitem{khalid_thesis24a}
\bibinfo{author}{Khalid, S.}
\newblock \emph{\bibinfo{title}{Quantum Optics as a theory of Point
  Processes}}.
\newblock Ph.D. thesis, \bibinfo{school}{University of Wolverhampton}
  (\bibinfo{year}{2024}).

\bibitem{schweickert18a}
\bibinfo{author}{Schweickert, L.} \emph{et~al.}
\newblock \bibinfo{title}{On-demand generation of background-free single
  photons from a solid-state source}.
\newblock \emph{\bibinfo{journal}{Appl. Phys. Lett.}}
  \textbf{\bibinfo{volume}{112}}, \bibinfo{pages}{093106}
  (\bibinfo{year}{2018}).
\newblock \urlprefix\url{doi:10.1063/1.5020038}.

\bibitem{hanschke18a}
\bibinfo{author}{Hanschke, L.} \emph{et~al.}
\newblock \bibinfo{title}{Quantum dot single-photon sources with ultra-low
  multi-photon probability}.
\newblock \emph{\bibinfo{journal}{npj Quantum Information}}
  \textbf{\bibinfo{volume}{4}}, \bibinfo{pages}{43} (\bibinfo{year}{2018}).
\newblock \urlprefix\url{doi:10.1038/s41534-018-0092-0}.

\bibitem{huber20a}
\bibinfo{author}{Huber, T.} \emph{et~al.}
\newblock \bibinfo{title}{Filter-free single-photon quantum dot resonance
  fluorescence in an integrated cavity-waveguide device}.
\newblock \emph{\bibinfo{journal}{Optica}} \textbf{\bibinfo{volume}{7}},
  \bibinfo{pages}{380} (\bibinfo{year}{2020}).
\newblock \urlprefix\url{doi:10.1364/OPTICA.382273}.

\bibitem{selvin75b}
\bibinfo{author}{Selvin, S.}
\newblock \bibinfo{title}{On the {Monty Hall} problem}.
\newblock \emph{\bibinfo{journal}{American Statistician}}
  \textbf{\bibinfo{volume}{29}}, \bibinfo{pages}{134} (\bibinfo{year}{1975}).
\newblock \urlprefix\url{https://www.jstor.org/stable/2683443}.

\bibitem{vossavant90a}
\bibinfo{author}{{vos Savant}, M.}
\newblock \bibinfo{title}{Ask {Marilyn}}.
\newblock \emph{\bibinfo{journal}{Parade Magazine}}
  \textbf{\bibinfo{volume}{September 9}}, \bibinfo{pages}{15}
  (\bibinfo{year}{1990}).

\bibitem{vazsonyi99a}
\bibinfo{author}{Vazsonyi, A.}
\newblock \bibinfo{title}{Which door has the {Cadillac}?}
\newblock \emph{\bibinfo{journal}{Decision Line}} \bibinfo{pages}{17}
  (\bibinfo{year}{1999}).

\bibitem{zubizarretacasalengua23a}
\bibinfo{author}{{Zubizarreta Casalengua}, E.}, \bibinfo{author}{del Valle, E.}
  \& \bibinfo{author}{Laussy, F.~P.}
\newblock \bibinfo{title}{Two-photon correlations in detuned resonance
  fluorescence}.
\newblock \emph{\bibinfo{journal}{Phys. Scr.}} \textbf{\bibinfo{volume}{98}},
  \bibinfo{pages}{055104} (\bibinfo{year}{2023}).
\newblock \urlprefix\url{doi:10.1088/1402-4896/acc89e}.

\bibitem{khalid_arXiv23a}
\bibinfo{author}{Khalid, S.} \& \bibinfo{author}{Laussy, F.}
\newblock \bibinfo{title}{Perfect single-photon sources}.
\newblock \emph{\bibinfo{journal}{arXiv:2306.13646}}  (\bibinfo{year}{2023}).

\bibitem{reynaud83a}
\bibinfo{author}{Reynaud, S.}
\newblock \bibinfo{title}{La fluorescence de r\'esonance: \'etude par la
  m\'ethode de l'atome habill\'e}.
\newblock \emph{\bibinfo{journal}{Annales de Physique}}
  \textbf{\bibinfo{volume}{8}}, \bibinfo{pages}{315} (\bibinfo{year}{1983}).
\newblock \urlprefix\url{http://cat.inist.fr/?aModele=afficheN&cpsidt=9386653}.

\bibitem{kim87a}
\bibinfo{author}{Kim, M.}, \bibinfo{author}{Knight, P.} \&
  \bibinfo{author}{Wodkjewicz, K.}
\newblock \bibinfo{title}{Correlations between successively emitted photons in
  resonance fluorescence}.
\newblock \emph{\bibinfo{journal}{Opt. Commun.}} \textbf{\bibinfo{volume}{62}},
  \bibinfo{pages}{385} (\bibinfo{year}{1987}).
\newblock \urlprefix\url{doi:10.1016/0030-4018(87)90005-8}.

\bibitem{brunel99a}
\bibinfo{author}{Brunel, C.}, \bibinfo{author}{Lounis, B.},
  \bibinfo{author}{Tamarat, P.} \& \bibinfo{author}{Orrit, M.}
\newblock \bibinfo{title}{Triggered source of single photons based on
  controlled single molecule fluorescence}.
\newblock \emph{\bibinfo{journal}{Phys. Rev. Lett.}}
  \textbf{\bibinfo{volume}{83}}, \bibinfo{pages}{2722} (\bibinfo{year}{1999}).
\newblock \urlprefix\url{doi:10.1103/PhysRevLett.83.2722}.

\bibitem{michler00a}
\bibinfo{author}{Michler, P.} \emph{et~al.}
\newblock \bibinfo{title}{A quantum dot single-photon turnstile device}.
\newblock \emph{\bibinfo{journal}{Science}} \textbf{\bibinfo{volume}{290}},
  \bibinfo{pages}{2282} (\bibinfo{year}{2000}).
\newblock \urlprefix\url{doi:10.1126/science.290.5500.2282}.

\bibitem{santori01a}
\bibinfo{author}{Santori, C.}, \bibinfo{author}{Pelton, M.},
  \bibinfo{author}{Solomon, G.}, \bibinfo{author}{Dale, Y.} \&
  \bibinfo{author}{Yamamoto, Y.}
\newblock \bibinfo{title}{Triggered single photons from a quantum dot}.
\newblock \emph{\bibinfo{journal}{Phys. Rev. Lett.}}
  \textbf{\bibinfo{volume}{86}}, \bibinfo{pages}{1502} (\bibinfo{year}{2001}).
\newblock \urlprefix\url{doi:10.1103/PhysRevLett.86.1502}.

\bibitem{ding16a}
\bibinfo{author}{Ding, X.} \emph{et~al.}
\newblock \bibinfo{title}{On-demand single photons with high extraction
  efficiency and near-unity indistinguishability from a resonantly driven
  quantum dot in a micropillar}.
\newblock \emph{\bibinfo{journal}{Phys. Rev. Lett.}}
  \textbf{\bibinfo{volume}{116}}, \bibinfo{pages}{020401}
  (\bibinfo{year}{2016}).
\newblock \urlprefix\url{doi:10.1103/PhysRevLett.116.020401}.

\bibitem{somaschi16a}
\bibinfo{author}{Somaschi, N.} \emph{et~al.}
\newblock \bibinfo{title}{Near-optimal single-photon sources in the solid
  state}.
\newblock \emph{\bibinfo{journal}{Nature Photon.}}
  \textbf{\bibinfo{volume}{10}}, \bibinfo{pages}{340} (\bibinfo{year}{2016}).
\newblock \urlprefix\url{doi:10.1038/nphoton.2016.23}.

\bibitem{aharonovich16a}
\bibinfo{author}{Aharonovich, I.}, \bibinfo{author}{Englund, D.} \&
  \bibinfo{author}{Toth, M.}
\newblock \bibinfo{title}{Solid-state single-photon emitters}.
\newblock \emph{\bibinfo{journal}{Nature Photon.}}
  \textbf{\bibinfo{volume}{10}}, \bibinfo{pages}{631} (\bibinfo{year}{2016}).
\newblock \urlprefix\url{doi:10.1038/nphoton.2016.186}.

\bibitem{wang19c}
\bibinfo{author}{Wang, H.} \emph{et~al.}
\newblock \bibinfo{title}{Boson sampling with 20 input photons and a 60-mode
  interferometer in a $10^{14}$-dimensional {Hilbert} space}.
\newblock \emph{\bibinfo{journal}{Phys. Rev. Lett.}}
  \textbf{\bibinfo{volume}{125}}, \bibinfo{pages}{250503}
  (\bibinfo{year}{2019}).
\newblock \urlprefix\url{doi:10.1103/PhysRevLett.123.250503}.

\bibitem{lopezcarreno18a}
\bibinfo{author}{{L\'opez Carre\~no}, J.~C.}, \bibinfo{author}{del Valle, E.}
  \& \bibinfo{author}{Laussy, F.~P.}
\newblock \bibinfo{title}{Frequency-resolved {Monte Carlo}}.
\newblock \emph{\bibinfo{journal}{Sci. Rep.}} \textbf{\bibinfo{volume}{8}},
  \bibinfo{pages}{6975} (\bibinfo{year}{2018}).
\newblock \urlprefix\url{doi:10.1038/s41598-018-24975-y}.

\bibitem{aspect13b}
\bibinfo{author}{Aspect, A.} \& \bibinfo{author}{Grangier, P.}
\newblock \bibinfo{title}{The first single-photon sources}.
\newblock \emph{\bibinfo{journal}{Exp. Methods Phys. Sci.}}
  \textbf{\bibinfo{volume}{45}}, \bibinfo{pages}{315} (\bibinfo{year}{2013}).
\newblock \urlprefix\url{doi:10.1016/B978-0-12-387695-9.00010-X}.

\bibitem{lopezcarreno18b}
\bibinfo{author}{{L\'opez Carre\~no}, J.~C.}, \bibinfo{author}{Casalengua,
  E.~Z.}, \bibinfo{author}{Laussy, F.} \& \bibinfo{author}{del Valle, E.}
\newblock \bibinfo{title}{Joint subnatural-linewidth and single-photon emission
  from resonance fluorescence}.
\newblock \emph{\bibinfo{journal}{Quantum Sci. Technol.}}
  \textbf{\bibinfo{volume}{3}}, \bibinfo{pages}{045001} (\bibinfo{year}{2018}).
\newblock \urlprefix\url{doi:10.1088/2058-9565/aacfbe}.

\bibitem{hanschke20a}
\bibinfo{author}{Hanschke, L.} \emph{et~al.}
\newblock \bibinfo{title}{Origin of antibunching in resonance fluorescence}.
\newblock \emph{\bibinfo{journal}{Phys. Rev. Lett.}}
  \textbf{\bibinfo{volume}{125}}, \bibinfo{pages}{170402}
  (\bibinfo{year}{2020}).
\newblock \urlprefix\url{doi:10.1103/PhysRevLett.125.170402}.

\bibitem{zubizarretacasalengua20a}
\bibinfo{author}{{Zubizarreta Casalengua}, E.}, \bibinfo{author}{{L\'opez
  Carre\~no}, J.~C.}, \bibinfo{author}{Laussy, F.~P.} \& \bibinfo{author}{del
  Valle, E.}
\newblock \bibinfo{title}{Conventional and unconventional photon statistics}.
\newblock \emph{\bibinfo{journal}{Laser Photon. Rev.}}
  \textbf{\bibinfo{volume}{14}}, \bibinfo{pages}{1900279}
  (\bibinfo{year}{2020}).
\newblock \urlprefix\url{doi: 10.1002/lpor.201900279}.

\bibitem{meyerscott20a}
\bibinfo{author}{Meyer-Scott, E.}, \bibinfo{author}{Silberhorn, C.} \&
  \bibinfo{author}{Migdall, A.}
\newblock \bibinfo{title}{Single-photon sources: Approaching the ideal through
  multiplexing}.
\newblock \emph{\bibinfo{journal}{Rev. Sci. Instrum.}}
  \textbf{\bibinfo{volume}{91}}, \bibinfo{pages}{041101}
  (\bibinfo{year}{2020}).
\newblock \urlprefix\url{doi:10.1063/5.0003320}.

\bibitem{reimer19a}
\bibinfo{author}{Reimer, M.~E.} \& \bibinfo{author}{Cher, C.}
\newblock \bibinfo{title}{The quest for a perfect single-photon source}.
\newblock \emph{\bibinfo{journal}{Nature Photon.}}
  \textbf{\bibinfo{volume}{13}}, \bibinfo{pages}{734} (\bibinfo{year}{2019}).
\newblock \urlprefix\url{doi:10.1038/s41566-019-0544-x}.

\bibitem{thomas21a}
\bibinfo{author}{Thomas, S.} \& \bibinfo{author}{Senellart, P.}
\newblock \bibinfo{title}{The race for the ideal single-photon source is on}.
\newblock \emph{\bibinfo{journal}{Nature Nanotech.}}
  \textbf{\bibinfo{volume}{16}}, \bibinfo{pages}{367} (\bibinfo{year}{2021}).
\newblock \urlprefix\url{doi:10.1038/s41565-021-00851-1}.

\bibitem{kirkwood39a}
\bibinfo{author}{Kirkwood, J.~G.}
\newblock \bibinfo{title}{Molecular distribution in liquids}.
\newblock \emph{\bibinfo{journal}{J. Chem. Phys.}}
  \textbf{\bibinfo{volume}{7}}, \bibinfo{pages}{919} (\bibinfo{year}{1939}).
\newblock \urlprefix\url{doi:10.1063/1.1750344}.

\bibitem{zernike27a}
\bibinfo{author}{Zernike, F.} \& \bibinfo{author}{Prins, J.~A.}
\newblock \bibinfo{title}{Die {Beugung} von {R\"ontgenstrahlen} in
  {Fl\"ussigkeiten} als {Effekt} der {Molek\"ulanordnung}}.
\newblock \emph{\bibinfo{journal}{Z. Phys. A}} \textbf{\bibinfo{volume}{41}},
  \bibinfo{pages}{184} (\bibinfo{year}{1927}).
\newblock \urlprefix\url{doi:10.1007/BF01391926}.

\bibitem{frenkel_book46a}
\bibinfo{author}{Frenkel, Y.}
\newblock \emph{\bibinfo{title}{Kinetic Theory of Liquids}}
  (\bibinfo{publisher}{Clarendon Press, Oxford}, \bibinfo{year}{1946}).

\bibitem{salsburg53a}
\bibinfo{author}{Salsburg, Z.~W.}, \bibinfo{author}{Zwanzig, R.~W.} \&
  \bibinfo{author}{Kirkwood, J.~G.}
\newblock \bibinfo{title}{Molecular distribution functions in a
  one‐dimensional fluid}.
\newblock \emph{\bibinfo{journal}{J. Chem. Phys.}}
  \textbf{\bibinfo{volume}{21}}, \bibinfo{pages}{1098} (\bibinfo{year}{1953}).
\newblock \urlprefix\url{doi:10.1063/1.1699116}.

\bibitem{sells53a}
\bibinfo{author}{Sells, R.~L.}, \bibinfo{author}{Harris, C.~W.} \&
  \bibinfo{author}{Guth, E.}
\newblock \bibinfo{title}{The pair distribution function for a
  one‐dimensional gas}.
\newblock \emph{\bibinfo{journal}{J. Chem. Phys.}}
  \textbf{\bibinfo{volume}{21}}, \bibinfo{pages}{1422} (\bibinfo{year}{1953}).
\newblock \urlprefix\url{doi:10.1063/1.1699263}.

\bibitem{wertheim63a}
\bibinfo{author}{Wertheim, M.~S.}
\newblock \bibinfo{title}{Exact solution of the percus-yevick integral equation
  for hard spheres}.
\newblock \emph{\bibinfo{journal}{Phys. Rev. Lett.}}
  \textbf{\bibinfo{volume}{10}}, \bibinfo{pages}{321} (\bibinfo{year}{1963}).
\newblock \urlprefix\url{doi:10.1103/PhysRevLett.10.321}.

\end{thebibliography}

\end{document}